# Assembling sequences of DNA using an on-line algorithm based on DeBruijn graphs

Andres Felipe Zapata Palacio
Universidad Eafit
Colombia
azapat47@eafit.edu.co

Juan Manuel Ciro Restrepo
Universidad Eafit
Colombia
jcirore@eafit.edu.co

Mauricio Toro
Universidad Eafit
Colombia
mtorobe@eafit.edu.co

**ABSTRACT**

The problem of assembling DNA fragments starting from imperfect strings given by a sequencer, classified as NP-hard when trying to get perfect answers [4], has a huge importance in several fields, because of its relation with the possibility of detecting similarities between animals, dangerous pests in crops, and so on. Some of the algorithms and data structures that have been created to solve this problem are Needleman–Wunsch algorithm, De bruijn graphs and greedy algorithms working on overlaps graphs; these try to work out the problem from different approaches that give place to certain advantages and disadvantages to be discussed.

In this article we first expose a summary of the research done on already created solutions for the DNA assembly problem, to present later an on-line solution to the same matter, which, despite not considering mutations, would have the capacity of using only the necessary amount of readings to assemble an user specified amount of genes.

**Author Keywords**

DNA assembly, DeBruijn graphs, eulerian walks, on-line algorithms, complexity.

**ACM Classification Keywords**

Applied computing→ Life and medical sciences→Computational biology→Recognition of genes and regulatory elements;

**INTRODUCTION**

As many other sciences, the biology progress has been markedly accelerated by the use of computational and statistical analysis in many fields; one of them, the assembly of DNA sequences. This specific problem has undergone a lot of variation in its constraints, from the hand work at the very beginning to the length of the reads of real time devices for DNA sequencing.

These devices, however, do not sequence the DNA in a predictable way, because the reads length is not enough to cover the whole DNA and it is not currently possible to start one exactly after another. Algorithms for genome assembly are then playing a crucial role in order to get entire genes.

**PROBLEM**

Currently, the DNA sequencers turn DNA in strings within the alphabet {A, G, T, C}, that represents its Nitrogenous bases. However, because a sequencer can not read a whole DNA directly, the DNA of a single creature is given in many fragments, which must be matched and conjugated on-line, it is, while the sequencer is still working.

Because of that, the solution has to work with the data introduced in real-time by the sequencer, and produce results before the reading process is finished. (It would take too much time and memory to wait for the reading to be completed)

The objective of this work is then finding one solution to process the reads of a sequencer with such characteristics, finding the DNA strand of the examined living being and extracting the genes found in the partially assembled DNA sequence.

The program will finish when the sequencer stops reading, when the number of genes required are found or when the limit time of computation is spent.

**RELATED WORK**

**De Bruijn graph and Eulerian walks [7]**
One algorithm created to solve the genome assembly problem is based on traversing a Bruijn graph following an Eulerian path (a trail in a graph which visits every edge exactly once[9]).

A Bruijn graph is a directed graph that represents each k-mer (Substring of length k) of the genome with nodes containing its prefix and suffix (k-1)-mers and then establishing an edge A → B for every A and B where B has a prefix that is also a suffix of A, that is, there is an overlap between A and B.

After building such a graph the genome can be obtained by following the aforementioned Eulerian path of the graph.

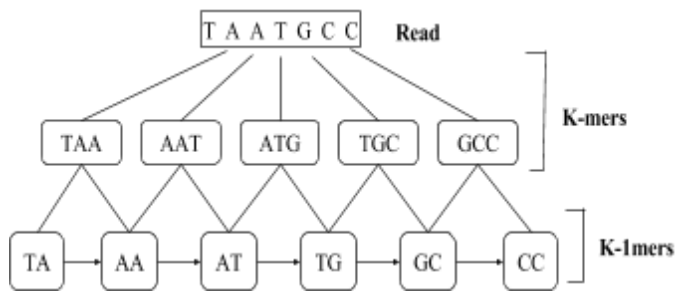

**Figure 1: graphic representation of a DeBruijn graph.**

**Global alignment of two genetic sequences [3]**
Needleman-Wunsch and Christian Wunsch implemented in 1970 an algorithm that aligns protein and nucleic acid sequences. This computation allows you to compare two sequences and determine how different or similar they are the one of the another.

This algorithm needs some inputs: the alphabet(set of symbols that make up the sequences), the two strings that contains the two sequences, a function that determines a mark of similarity between each pair of symbols that make up the alphabet, and the mark of no matching a symbol.

**Greedy shortest common superstring [4]**
Another approach to the genome assembly problem, simplified to finding the shortest common superstring, are greedy algorithms, which are quick at the cost of not guaranteeing that the found superstring is the shortest one. More precisely, the greedy approximations can compute in $O(n \log n)$ time a superstring that in the worst case is, however, "only $\beta$ times ( where $2 \leq \beta \leq 4$) longer than the shortest common superstring"[6].

The strategy behind the greedy approach is basically building and "greedily" reducing an overlaps graph. This graph is a directed weighted graph, where $A \rightarrow B$ shows that A overlaps B, and the weight of the edge indicates the number of overlapped characters. The reduction, on the other hand, merges (eliminating redundancy) the nodes attached to the edge with the greatest weight at each iteration, randomly choosing if there is more than one edge labeled with that weight and concatenating the remaining nodes after all edges have been merged.

**Comparing biological sequence information [5]**
BLAST is an informatic program that computes the statistical level of similarity between a given nucleotide sequence and all the sequences stored in its database, using an heuristic algorithm.

Although the heuristic does not ensures the answer is correct, in most of the cases, BLAST works not only successfully, but also efficiently.

## ALGORITHM

This algorithm uses one DeBruijn graph as in the Figure 1, to store the k-mers of each fragment read:

The program reads each DNA fragment given and updates the DeBruijn graph:, getting the right and left K-1mers for each kmer, inserting them into the graph (if they are not already added) and then setting a directed edge from the left to the right one. Also, it is necessary to discard the possibility that the right K1-mer (the one to which the edge "arrives") is initial, this aspect is going to be clarified later.

Once the graph is updated, the next step is to traverse it. It is done by following the next K-1mers of each initial K-1mer (each K-1mer to which no other K-1mer points), until a K-1mer without next is found. This process gave us a set of partially assembled strands.

The last step is processing every strand and finding the genes in it. At first the program goes over the strand and saves the indexes of the start and stop codons, after that it "greedily" matches the first start codon found with the first stop codon (a gen has been found) and the start codons in between are removed, that because there can be many start codons before a stop codon and the "correct" one is the one that produces the largest gen.

If one gen is found, the program checks if it is already found(the found genes are stored in a HashSet so that it is easy to check) if that is not the case, the gen is printed in the screen and stored in that HashSet.

## DATA STRUCTURES

The needings that the data-structure-level graph representation had to suffice are basically three: Each K-1 mer needs to be added at most once, the graph must be able to be efficiently traversed, and, as a consequence of the former, the initial nodes should be able to be found easily.

At a first glance, a HashMap<K-1mer, nextK-1mer> seems to be a good option: It is $O(1)$ for inserting, and also allows to get the "next" of each K-1mer in constant time, but when it comes to finding the initial nodes there is no straightforward way to do that.

Keeping in mind the HashMap advantages we implemented a bijection between each K-1mer and an index using a bidirectional HashMap, so that getting the index of a K-1mer and the K-1mer of an index are both performed in $O(1)$ constant time. That indexing makes it possible to represent the graph as an one-dimensional "adjacency array" where every edge $A \rightarrow B$ is represented as array[indexof A] = indexof B. The array is one-dimensional since A and B are guaranteed to be unique, which causes that edge to be unique, and the only outgoing edge from A.

To find the initial nodes the the strategy is pretty simple: every K-1mer, at the beginning, is a "possible initial", and each time an edge A → B is added, B stops being an initial. We save that information in a 1D boolean array, indexed the same way, and which values are all True at the beginning. In the end those K-1mers whose values in their corresponding indices are still True are the initial nodes:

Let A, B and C be K-1mers with indices 0, 1 and 2 respectively. All of them are initials at the beginning:

| T | T | T |
|---|---|---|
| 0 | 1 | 2 |

After reading AB the arc A→B is created. Then, B is not a possible initial anymore:

| T | F | T |
|---|---|---|
| 0 | 1 | 2 |

**Figure 2: Strategy used to determine the initial nodes.**

This implementation assures us that the same edge or K-1mer will not be stored twice. Shown below, there is a summary of the complexity of the operations gotten by using this implementation.

**V: number of K-1mers stored in the graph**

| Operation | Complexity |
|---|---|
| Get the index of one K-1mer | O(1) |
| Get the K-1mer referred by one index | O(1) |
| Get the next K-1mer of one K-1mer | O(1) |
| Check whether a K-1mer is an initial node | O(1) |
| Get all the initial nodes of the graph | O(V) |
| Traverse the graph(starting at every initial node) | O(V) |

**Table 1: Complexity of operations on the proposed implementation of the graph.**

**COMPLEXITY**

Since the length of the K-mer is 201, the graph is going to store K1-mers of 200. When the graph is being updated, for each K-mer in the read of length L, two K-1mers are taken, that is $(L-201)*2$. After that, theY new K-1 mers are inserted and their edges added to the graph, both operations in constant time. **Updating the graph is then $O(L)$.**

Once the graph is updated it has to be traversed in order to find the genes. First, we get all the initials O(V), and then traverse the graph from every initial. Since the group of K-1mers pointed by each initial are disjoint every K-1mer is visited only once and then **the complexity of traversing the graph is** $O(V+V) = O(2V) = O(V)$

Putting it all together, the complexity of processing **N** lectures is developed next:

For the sake of simplicity we use the average read length instead of the specific lengths of all reads. It is fair considering that they were going to be used altogether in a summatory and, because of that, the final value does not change.

Let A be the average read length. Each time the graph is traversed every already-added node is visited, that would be A for the first read, plus A + A for the second one and so on. In the nth lecture, the amount of traveled nodes would be:

$$A + 2A + 3A + ... + NA$$

Factoring A we get:

$$(1 + 2 + ... + N) * A$$

Which, using Gauss summation, can be reduced to:

$$\frac{N(N+1)}{2} * A = \frac{AN^2 + AN}{2}$$

Now, applying Big O product, constant and sum rules we get:

$$O(\frac{AN^2 + AN}{2}) = O(AN^2 + AN) = O(AN^2)$$

That would be the complexity of traversing the graph N times. The complexity of updating the graph is way more simple: it walks each read completely, which is O(AN).

The overall complexity, adding the cost of traversing and updating the graph is $O(AN + AN^2)$, and, using the product rule once again, **we conclude that the complexity of processing N reads is $O(AN^2)$ where A is the average read length.**

| Operation | Complexity |
|---|---|
| Update the graph with a read of length L | O(L) |
| Traverse the graph when it has V K-1mers added | O(V) |
| Update and traverse the graph N times, with N reads of average length A | O(AN²) |

**Table 2: Complexity of the operations used by the algorithm.**

## IMPLEMENTATION

The program execution requires three arguments: the filename where the DNA fragments are stored, the number of genes to be found and the computing timeout.

```
Usage: java DNAReader <file> <nGenes> <timeLimit> [ -c ]
<file>      name of the file that contains the readings
<nGenes>    number of genes to be found
<timeLimit> maximum miliseconds of computation time
-c          allows the user to insert manually the start and stop codons
```

**Figure 3:** Help menu when the arguments are incorrects

A fourth parameter is optional, represented by the flag -c; this argument allows the user to write the start and stop codons, separated by commas.

```
$java DNAreader2 Segments/AcipenserTransmontanus.txt 1 3 -c
Enter the initial codons separated by spaces:
atg
Enter the final codons separated by spaces:
tga taa
```

**Figure 4:** Starting the program with the -c argument

If the flag -c is not written, the default start codon is ATG and the default stop codons are TGA, TAA and TAG.

During the program execution, the found genes are printed in the screen one by one.

```
atgttaagatgagccctagacagctccgcaggcacaaaggcttggtcctggccttactatcaattttaac
ttgctaagccacaccccaagggaactcagcagtgataaacattga
atgagcgcaagctcgactcagccagagttaagagggccggtaaaactcgtgccagccaccgcggttatac
aaccacgaaggtagctctacctaacaaggacccctttgaacccacgacaactgagacacaaactgggatta
Found 2 fragments in 12.0ms
```

**Figure 5:** Final results of the execution

At the end of the execution, the program prints the number of found fragments and the computing time expressed in milliseconds.

The code can be found at:
https://svn.riouxsvn.com/edya-dnassembly

## RESULTS WITH TWO SMALL DNAS

The implementation was tested at assembling mitochondrial genes from two species: Acipenser transmontanus (White sturgeon) and Acanthisitta chloris (rifleman). The memory and time results are shown below:

| Species: Acipenser transmontanus | | | | |
|---|---|---|---|---|
| Number of requested genes | 15 | 30 | 45 | 60 |
| Execution time(ms) | 38 | 70 | 132 | 284 |
| Memory(MB) | 23,2 | 29,4 | 35,3 | 62,8 |

**Table 3:** Time and memory results from the tests with mitochondrial Acipenser transmontanus DNA

| Species: Acanthisitta chloris | | | | |
|---|---|---|---|---|
| Number of requested genes | 15 | 30 | 45 | 60 |
| Execution time(ms) | 53 | 126 | 212 | 249 |
| Memory(MB) | 25,9 | 42,8 | 60,8 | 62,6 |

**Table 4:** Time and memory results from the tests with mitochondrial Acanthisitta chloris DNA

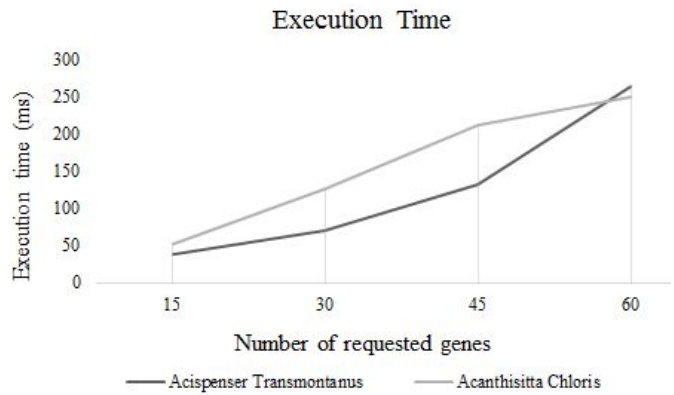

**Figure 6:** Results of time for both tests.

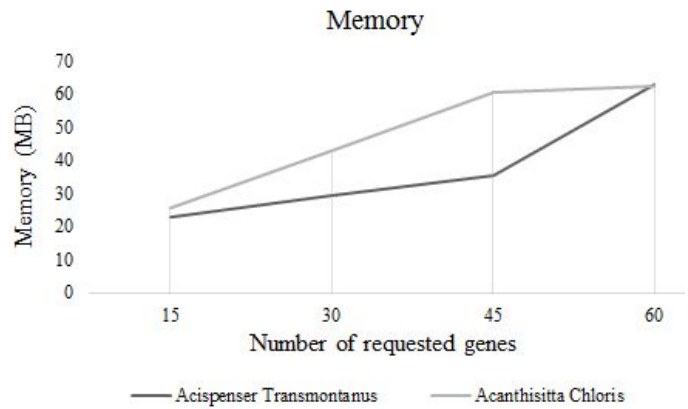

**Figure 7:** Memory used by both tests.

## CONCLUSIONS

The problem of assembling DNA sequences from a group of fragments using an on-line algorithm is a wide problem for which every assumption, made or not, can influence in high degree the difficulty or ease on the solution. This project's difficulty was increased by the fact that the algorithm had to be on-line, but also reduced because mutations were not considered.

The results of time and memory were totally satisfactory and it is certain that the software performs well assembling genes from DNAs under, or slightly above 20000 nucleobases long. We also conclude that it is impossible to assembly genes with 100% of correctness by using on-line algorithms, specially because there is no way to make predictions about start or stop codons that are yet to come in further reads.

Thanks to this project, we got started in the world of scientific computing, using our programming knowledge for solving a computational genomic problem: assembling a DNA sequence and finding the genes using an on-line algorithm.

This approach to the world of science shows us the large number of problems where the use of computing can make a difference. Science and computing are nowadays meant to walk together in order to find answers and solutions for the benefit of mankind.

## ACKNOWLEDGEMENT

Special thanks to Ben Langmead from the John Hopkins School of engineering, whose courses and online-sources gave us the theoretical framework for this project.

Also to Santiago Passos Patiño and Juan David Arcila from EAFIT university, for the space to discuss different approaches to solve the proposed problem.

## FUTURE WORK

The future works could be focused on reducing the computation time, because we are still solving some subproblems more than once. Another important improvement would be modifying the algorithm in order to consider mutations and the statistical models behind this.